\documentclass[twoside,twocolumn,9pt]{article}
\usepackage{extsizes}

\usepackage[super,sort&compress,comma]{natbib}
\usepackage[version=4]{mhchem}
\usepackage[left=1.5cm, right=1.5cm, top=1.785cm, bottom=2.0cm]{geometry}
\usepackage{balance}
\usepackage{mathptmx}
\usepackage{sectsty}
\usepackage{graphicx} 
\usepackage{lastpage}
\usepackage[format=plain,justification=justified,singlelinecheck=false,font={stretch=1.125,small,sf},labelfont=bf,labelsep=space]{caption}
\usepackage{float}
\usepackage{fancyhdr}
\usepackage{fnpos}
\usepackage[english]{babel}
\addto{\captionsenglish}{%
  
}
\usepackage{array}
\usepackage{droidsans}
\usepackage{charter}
\usepackage[T1]{fontenc}
\usepackage[usenames,dvipsnames]{xcolor}
\usepackage{setspace}
\usepackage[compact]{titlesec}
\usepackage{hyperref}

\usepackage{epstopdf}

\definecolor{cream}{RGB}{222,217,201}

\usepackage{booktabs}       
\usepackage[list-units=single,detect-all]{siunitx}
\DeclareSIUnit{\calorie}{cal}
\DeclareSIUnit{\kcal}{\kilo\calorie\per\mol}
\DeclareSIUnit{\angstrom}{\text {Å}}

\begin{document}

\makeFNbottom
\makeatletter
\renewcommand\LARGE{\@setfontsize\LARGE{15pt}{17}}
\renewcommand\Large{\@setfontsize\Large{12pt}{14}}
\renewcommand\large{\@setfontsize\large{10pt}{12}}
\renewcommand\footnotesize{\@setfontsize\footnotesize{7pt}{10}}
\makeatother

\renewcommand{\thefootnote}{\fnsymbol{footnote}}
\renewcommand\footnoterule{\vspace*{1pt}%
\color{cream}\hrule width 3.5in height 0.4pt \color{black}\vspace*{5pt}} 
\setcounter{secnumdepth}{5}

\makeatletter 
\renewcommand\@biblabel[1]{#1}            
\renewcommand\@makefntext[1]%
{\noindent\makebox[0pt][r]{\@thefnmark\,}#1}
\makeatother 
\renewcommand{\figurename}{\small{Fig.}~}
\sectionfont{\sffamily\Large}
\subsectionfont{\normalsize}
\subsubsectionfont{\bf}
\setstretch{1.125} 
\setlength{\skip\footins}{0.8cm}
\setlength{\footnotesep}{0.25cm}
\setlength{\jot}{10pt}
\titlespacing*{\section}{0pt}{4pt}{4pt}
\titlespacing*{\subsection}{0pt}{15pt}{1pt}

\newcommand{\etal}{\textit{et al.}}

\twocolumn[
  \begin{@twocolumnfalse}

\begin{center}
\LARGE{\textbf{Geometries, interaction energies and bonding in \ce{[Po(H2O)_n]^{4+}} and \ce{[PoCl_n]^{4-n}} complexes$^\dag$}}
\end{center}

\vspace{0.3cm}

\begin{center}
\large{Nadiya Zhutova,\textit{$^{a,b}$} Florent Réal,\textit{$^{c}$} Valérie Vallet,$^{\ast}$\textit{$^{c}$} and Rémi Maurice$^{\ast}$\textit{$^{a,b}$}}
\end{center}

\vspace{0.3cm}

\noindent\normalsize{Polonium ($Z$ = 84) is one of the rarest elements on Earth. More than a century after its discovery, its chemistry remains poorly known and even basic questions are not yet satisfactorily addressed. In this work, we perform a systematic study of the geometries, interactions energies and bonding in basic polonium(IV) species, namely the hydrated \ce{[Po(H2O)_n]^{4+}} and chlorinated \ce{[PoCl_n]^{4-n}} complexes by means of gas-phase electronic structure calculations. We show that while up to nine water molecules can fit in the first coordination sphere of the polonium(IV) ion, its coordination sphere can already be filled with eight chloride ligands. Capitalising on previous theoretical studies, a focused methodological study based on interaction energies and bond distances allows us to validate the MP2/def2-TZVP level of theory for future ground-state studies. After discussing similarities and differences between complexes with the same number of ligands, we perform topological analyses of the MP2 electron densities in the quantum theory of atoms in molecules (QTAIM) fashion. While the water complexes display typical signatures of closed-shell interactions, we reveal large Po--Cl delocalisation indices, especially in the hypothetical \ce{[PoCl]^{3+}} complex. This ``enhanced'' covalency opens the way for a significant spin-orbit coupling (SOC) effect on the corresponding bond distance, which has been studied by two independent approaches (\textit{i.e.} one \textit{a priori} and one \textit{a posteriori}). We finally conclude by stressing that while the SOC may not affect much the geometries of high-coordinated polonium(IV) complexes, it should definitely not be neglected in the case of low-coordinated ones.} 

 \end{@twocolumnfalse} \vspace{0.3cm}

\vspace{0.6cm}

]

\footnotetext{\textit{$^{a}$~Subatech, UMR CNRS 6457, IN2P3/IMT Atlantique/Universit\'e de Nantes, 4 rue A. Kastler, 44307 Nantes Cedex 3, France}}
\footnotetext{\textit{$^{b}$~Univ Rennes, CNRS, ISCR (Institut des Sciences Chimiques de Rennes) -- UMR 6226, F-35000 Rennes, France; E-mail: remi.maurice@univ-rennes1.fr}}
\footnotetext{\textit{$^{c}$~Univ. Lille, CNRS, UMR 8523-PhLAM-Physique des Lasers, Atomes et Molécules, F-59000 Lille, France; E-mail: valerie.vallet@univ-lille.fr}}
\footnotetext{\dag~Electronic Supplementary Information (ESI) available: Additional interaction energies, bond distances and atomic charges obtained at various levels of theory (PDF) and MP2/def2-TZVP molecular geometries for all the studied complexes (XYZ files compressed into two ZIP files). }

\section{Introduction}

Since the discovery of polonium in 1898 by Marie and Pierre Curie, not a lot of experiments has been carried out to unravel its intriguing chemical properties. While most of polonium chemistry remains unknown, its extreme toxicity, being it either chemical or radiological, is notorious~\cite{Ansoborlo:2012, Ansoborlo:2014}. Though its radioactivity may be used to detect it, it is still complicated to perform analyses to determine polonium traces in, for example, sea water~\cite{shannon1970polonium}, tobacco~\cite{khater2004polonium, radford1964polonium}, plants~\cite{wieczorek2022determination}, soils~\cite{Le:2019} and other samples. Meanwhile, former usages of polonium as a heat source in the space machine industry, as a $\alpha$-emission source, or as static eliminators have now become obsolete because of alternatives based on materials that are safer and easier to get. In the case of accidental or criminal ingestion of polonium, no specific antidote is known, and one can administrate British anti-Lewisite (BAL), the antidote for heavy metals and in particular arsenic~\cite{Vilensky:2003}, unfortunately without any guarantee of success.  

Polonium has more than 40 known isotopes. While only seven of them are naturally occurring~\cite{brown2019aqueous}, \ce{^{210}Po} is the usual isotope for performing experiments, even if other ones may be scarcely used, \textit{e.g.} \ce{^{208}Po} for atomic spectroscopy~\cite{Charles:1966}.  Four most oxidation states of polonium are usually considered: $-$II, +II, +IV and +VI. In solution, polonium(IV) is meant to be the most stable one~\cite{figgins1961radiochemistry}, so that it is clearly an oxidation state whose chemistry deserves to be studied. Nonetheless, evidences for polonides (\ce{Na2Po}, \ce{ZnPo}, \textit{etc.})~\cite{moyer1956polonium}, oxides (\ce{PoO2}, \ce{PoO3})~\cite{bagnall1954preparation}, halides and dihalides (\ce{PoCl2}, \ce{PoCl4}, \ce{PoBr4}, \ce{PoI4})~\cite{bagnall1955poloniumCl,bagnall1955poloniumBr,bagnall1956657} and complexes with organic ligands such as the acetylacetonate anion (acac$^-$), the ethylenediaminetetraacetic acid trianion (\ce{EDTA^{3-}}), the oxalate dianion (\ce{C2O4^{2-}}), \textit{etc.} were reported in the literature~\cite{ampelogova1973}.  

Computational chemistry is, of course, free of any direct toxicological impact. Its increased accuracy and applicability observed in the last few decades may open the way for future predictions aiming at designing new successful experiments with polonium. However, prior to this, methodological studies have to be performed in order to ensure good results for the good reasons (or at least identified error cancellations). In fact, only a limited number of electronic structure theory studies were dedicated to polonium, among which a few emblematic ones will be discussed hereafter. Note that the electronic structure of the free \ce{Po^{4+}} ion is [Xe]4f$^{14}$5d$^{10}$6s$^2$, which has two main consequences for the calculations, (i) single-configurational and single-reference approaches may correctly describe the ground state wave functions of polonium(IV) species and (ii) relativistic effects must be accounted for explicitly or implicitly (\textit{e.g.} scalar relativistic effects lead to a strong stabilisation of the 6s orbital~\cite{Thayer:2005a}). 

From 2008 to 2010, Ayala~{\etal} published three computational works related to polonium(IV) hydration and to the formation of hydroxo complexes in solution~\cite{ayala2008po,ayala2009general,ayala2010ab}. In 2015, Borschevsky~{\etal} calculated very accurately with a fully relativistic coupled cluster method the ionisation potentials and electron affinity of the free Po atom~\cite{borschevsky2015ionization}. Note that their first ionization potential value was in good accord with the early value of Charles~\cite{Charles:1966}, and proved to be also in good accord with the yet to come new experimental values derived from two independent experiments~\cite{fink2019determination,raeder2019determination}. In 2019, Stoïanov~{\etal} revisited from theory the UV-Vis absorption spectra of polonium chloride complexes~\cite{stoianov2019uv}. The same year, chalcogen bonding with polonium was examined by Zierkiewicz~{\etal}~\cite{zierkiewicz2019chalcogen}. In fact, performing relativistic and correlated calculations is always quite challenging, and validating approximate schemes from scratch may require tedious methodological studies due to the lack of experimental reference data (one should then validate more approximate schemes by comparison with more accurate ones, \textit{i.e.} based on very demanding calculations, see for instance the work of Rota \textit{et al.} on the polonium hypothetical dimer~\cite{rota2011zero}).

As mentioned before, the +IV oxidation state of polonium dominates in solution. Following the pioneer works of Ayala~{\etal}~\cite{ayala2008po} and Sto\"ianov~{\etal}~\cite{stoianov2019uv}, we will focus on two families of Po(IV)-complexes, namely the hydrated \ce{[Po(H2O)_n]^{4+}} and the chlorinated \ce{[PoCl_n]^{4-n}} complexes. The first series is of fundamental interest since it relates to the general context of the hydration of heavy-element cations in solution~\cite{ram2019aqueous}. The second series is of practical interest since polonium chloride complexes can form in acidic chloride solutions~\cite{moyer1956polonium,ram2019aqueous}. We will focus on ground-state properties (geometries, interaction energies and bonding descriptors) and assess the dependence of the theoretical results with respect to specific degrees of freedom such as the choice of the correlation method, of the atomic basis sets or for instance of correcting for the basis set superposition error (BSSE) when computing interaction energies, which was not fully done in those earlier works~\cite{ayala2008po,stoianov2019uv}. To limit the discussion to pure electronic-structure theory aspects, only gas-phase results will be considered (future work may deal with solvation). Similarities and differences between the hydrated and chlorinated complexes will be discussed, in particular concerning the first coordination sphere of the polonium(IV) cation. 

\section{Computational details}

Our most extended methodological investigation was done on the \ce{[Po(H2O)_n]^{4+}} complexes (where in fact $n$ = 1--9), even though some tests were repeated on the \ce{[PoCl_n]^{4-n}} ones ($n$ = 1--8, see Results and discussion section and Supporting information). MP2~\cite{frisch1990direct,frisch1990semi,head1988mp2,saebo1989avoiding,head1994analytic} and B3LYP~\cite{lee1988development,raghavachari2000perspective} calculations were performed to determine the molecular geometries and the corresponding total interaction energies of all the complexes. Because of the close agreement of the MP2 and B3LYP results and for the sake of simplicity, we will only report MP2 results in the remainder of the text. For the \ce{[Po(H2O)_n]^{4+}} ($n$ = 7--9) complexes, initial structures were built from previous thorium structures~\cite{real2010quantum}. Note that additional reference calculations with the coupled cluster with single, double and perturbative triple excitations~\cite{purvis1982full,pople1987quadratic}, CCSD(T), method were also performed on specific complexes, namely the \ce{[Po(H2O)n]^{4+}} with $n$ = 1--3. All the MP2, B3LYP and CCSD(T) calculations were carried out with the Gaussian 16~\cite{g16} software. Note that every molecular geometry is optimised unless mentioned otherwise. 

The Dunning~\cite{peterson2003systematically} aug-cc-pVDZ, aug-cc-pVTZ, and aug-cc-pVQZ, and redefined Ahlrichs~\cite{defBS} def2-TZVP and def2-QZVP basis sets together with the relativistic 60-electron small core ECP60MDF pseudopotential~\cite{peterson2003systematically} for Po and its associated aug-cc-pV$N$Z-PP or def2-$N$ZVP-PP basis set were used ($N$ =  D, T, Q). All the basis sets and pseudopotential parameters were taken from the Basis Set Exchange software~\cite{pritchard2019a}. 

\begin{figure}[ht]
\centering
  \includegraphics[scale=0.4]{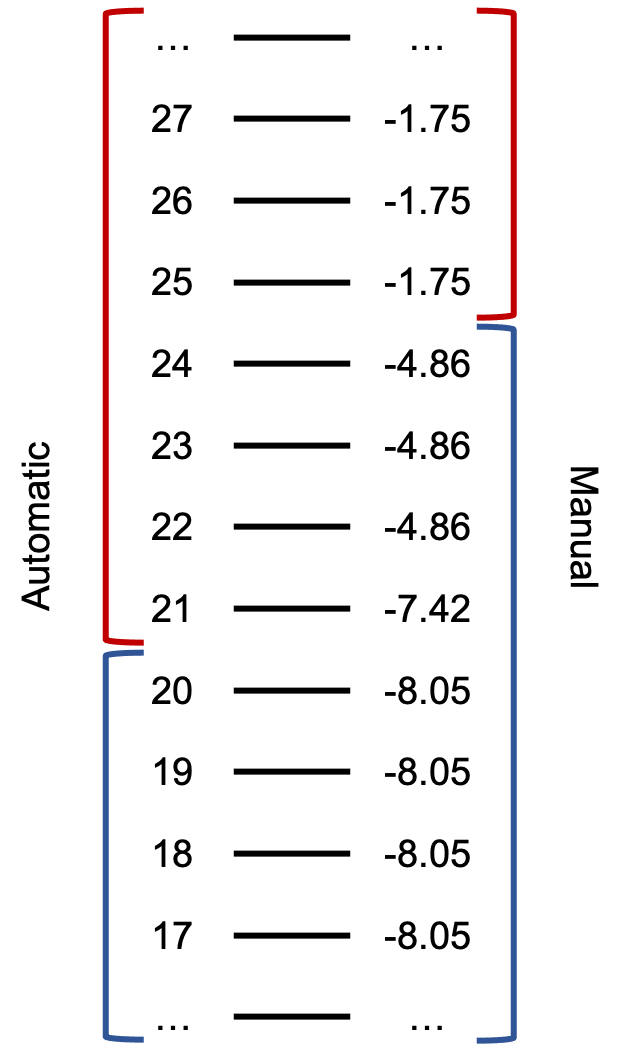}
  \caption{Hartree-Fock/def2-TZVP orbital energies (in a.u.) of the \ce{PoCl4} complex. The frozen (blue) and unfrozen (red) orbitals are displayed, by default (left) or with our recommended scheme (right).}
  \label{fgr:NFC}
\end{figure}

By default, in post-Hartree-Fock calculations, Gaussian~16 correlates all outer-core (5s, 5p, 5d) orbitals and the occupied 6s valence orbital of Po(IV), while freezing the 5s and 5p orbitals is wiser to ensure a significant energy gap between the frozen and the correlated orbitals in a molecular system, as illustrated by Figure~\ref{fgr:NFC} related to the \ce{PoCl4} molecule, where the Po 5s  molecular orbital (MO 21) is energetically close to the frozen combinations of the chlorine 2s (MOs 9--12) and 2p orbitals (MOs 13--20). To generally avoid artefacts in the correlated calculations (when discussing trends, one should only compare results for which the same shells are frozen), the default frozen core choice was overwritten with the $window$ keyword of Gaussian, with which the user may define the first unfrozen orbital, \textit{i.e.} the number of frozen cores --NFC-- plus one). In fact, in the current study, it was manually set to freeze 4 orbitals for Po (5s and 5p pseudo-orbitals), 1 for each water (the 1s orbital of the O atom), and 5 for Cl (the 1s, 2s and 2p shells). Moreover, though it is a bit technical, we would like to warn the reader that attention should be paid in computing the frequencies in a second step to ensure that Gaussian correctly applies the desired freezing scheme. 

The interaction energies were calculated using the following formula:
\begin{equation}
\label{Eq:Eint}
  E_{int} = E_{complex}-E_{Po^{4+}}-nE_{ligand}+\Delta E_\text{BSSE}
\end{equation} 

\noindent The basis set superposition error correction ($\Delta E_\text{BSSE}$) was calculated by the means of the counterpoise correction at the (fixed) geometry of the complex:
\begin{equation}
\label{Eq:EBSSE}
  \Delta E_\text{BSSE} = E_{\text{Po}^{4+}}-E_{\text{Po}^{4+}}^\text{ghosts}-E_{(\text{ligands})_n}+E_{(\text{ligands})_n}^\text{ghost}
\end{equation} 

\noindent where the \emph{ghost(s)} superscript refers to calculations performed with ghost ligand atoms ($E_{\text{Po}^{4+}}^\text{ghosts}$) or with a ghost Po atom ($E_{(\text{ligands})_n}^\text{ghost}$), meaning that the basis functions of the corresponding  atoms are applied but their electrons are not (neither their nuclear charges). Note that unlike  Equation \ref{Eq:Eint}, which dissociates the complex in \ce{Po^{4+}} + $n$ ligands, Equation \ref{Eq:EBSSE} considers clusters of $n$ interacting ligands, denoted $(\text{ligands})_n$.

To describe the nature of the chemical bonds in the systems under study, quantum theory of atoms in molecules (QTAIM)~\cite{Bader:1991} analyses were done with the AIMALL~\cite{aimall191012} software using the MP2/def2-TZVP electron densities. Note that more details concerning the meaning of the derived descriptors will be directly given in the Results and discussion section. 

To assess the role of the spin-orbit coupling (SOC) on molecular geometries, two types of calculations were performed. First, zeroth-order regular approximation (ZORA)~\cite{vanLenthe:1993} relativistic Hartree-Fock calculations were performed using the AMS driver of the Amsterdam Modeling Suite~\cite{ADF2001} software and Slater-type TZ2P~\cite{van2003optimized} basis sets, without freezing core orbitals. In this case, the SOC is included \textit{a priori}, but the interplay between SOC and electron correlation is completely missing. Alternatively, we have performed two-step SOC calculations with ORCA~\cite{Neese:2022} (v. 5.0.0), for which the SOC is introduced \textit{a posteriori}, as a perturbation of the correlated spin-orbit free (SOF) level, obtained with the Douglas-Kroll-Hess Hamiltonian~\cite{Douglas:1974,Hess:1986,Jansen:1989}. In this case, specific scalar relativistic recontracted (DKH-def2-TZVP) or originally generated (SARC-DKH-TZVP) all-electron basis sets~\cite{Pantazis:2008,pantazis2012all} were used for all the atoms. There, the role of the SOC may this time be underestimated, and only a part of the interplay between electron correlation and the SOC is by construction treated (we have only used the contracted spin-orbit configuration interaction, SOCI, scheme). More details and explanation will be directly given in the Results and discussion section. In short, these two complementary approaches do not suffer from the same pitfalls, and if a physical effect is observed with both methodologies, one may thus be quite confident in its actual occurrence. 

\section{Results and discussion}

\subsection{Electronic structure theory calculations}

As mentioned earlier, we will start the discussion with the interactions energies that were obtained for the \ce{[Po(H2O)_n]^{4+}} and \ce{[PoCl_n]^{4-n}} complexes, with special emphasis on the water complexes, previously studied by Ayala~{\etal}~\cite{ayala2008po}. We performed a partial benchmark to find the best accuracy/cost methodology for ground-state studies of polonium(IV) complexes. We recall here some of their important conclusions, confirmed by our independent calculations:

\begin{itemize}
    \item up to nine water molecules can fit in the polonium(IV) coordination sphere, meaning that $n$ = 1--9 for the hydrated \ce{[Po(H2O)_n]^{4+}} complexes under study;
    \item the interaction energies evolve monotonously with $n$;
    \item MP2 and to a lesser extend B3LYP calculations provide interactions energies in very good agreement with the CCSD(T) ones when the aug-cc-pVDZ basis set are used.
\end{itemize}
We have thus retained the MP2 method, and focused on two main degrees of freedom: the nature and size of the basis set and the use of a BSSE correction (see Eqs.~\ref{Eq:Eint} and \ref{Eq:EBSSE}). Furthermore, since we have proposed to change the number of polonium frozen orbitals (see Computational details section), we have also reported new CCSD(T) and MP2 calculations with the aug-cc-pVDZ basis set, as well as original ones with triple zeta and quadruple zeta aug-cc or def2 basis sets (see Table \ref{tbl:Eint}, as well as Tables S1 and S2 for an investigation of the impact of the BSSE correction). 

\begin{table*}[ht]
	\small
\caption{\ Interaction energies with BSSE correction (\si{\kcal}) obtained at various levels of theory for polonium(IV) complexes with water and chlorides, computed according to Eqs.~\ref{Eq:Eint} and \ref{Eq:EBSSE}.}
\label{tbl:Eint}
\begin{tabular*}{\textwidth}{@{\extracolsep{\fill}}l*{6}{S[table-format=-4.1]}}
	\toprule
	    & \multicolumn{5}{c}{\ce{H2O}} & \ce{Cl-}\\
	    \cmidrule{2-6}
		{$n$(ligands)} & {CCSD(T) /} & {MP2 /} & {MP2 /} & {MP2 /} & {MP2 /} & {MP2 /} \\
		& {aug-cc-pVDZ} & {aug-cc-pVDZ} & {aug-cc-pVTZ} & {def2-TZVP} & {def2-QZVP} & {def2-TZVP} \\
		\midrule
		1 & -236.2 & -249.2 & -257.9 & -262.0 & -263.8 & -883.7 \\
    2 & -409.5 & -439.0 & -454.7 & -461.4 & -464.9 & -1415.0 \\
    3 & -541.1 & -586.4 & -607.7 & -616.1 & -621.4 & -1749.2 \\
    4 &  & -694.9 & -718.8 & -728.9 &  & -1911.4 \\
    5 &  & -791.8 & -817.6 & -829.4 &  & -1974.3 \\
    6 &  & -875.8 & -903.1 & -918.2 &  & -1947.8 \\
    7 &  & -941.2 & -971.1 & -987.8 &  & -1861.4 \\
    8 &  & -1004.4 & -1035.9 & -1054.0 &  & -1650.3 \\
    9 &  & -1052.7 & -1085.5 & -1104.4 &  & \\
		\bottomrule
	\end{tabular*}
\end{table*}

One can quickly see that results for both the MP2 and CCSD(T) methods, in combination with the aug-cc-pVDZ basis set, are in fair agreement, though the accord is less impressive than in the paper of Ayala \textit{et al.}~\cite{ayala2008po}. Since the BSSE corrections are similar with both methods (see Tables S1 and S2), the observed difference essentially comes from the number of frozen orbitals. However, it is known that wave-function theory calculations, especially correlated ones of the coupled-cluster type, slowly converge with the basis-set size. Because of the computational cost, we have not systematically performed CCSD(T) calculations with larger basis sets and have essentially investigated the effect of the basis-set nature and size only at the MP2 level. However, the interested reader may find a couple of CCSD(T) optimised structures obtained with the aug-cc-pVTZ and def2-TZVP basis sets in Table S2. Interestingly, the accord between MP2 and CCSD(T) improves with basis set enlargement, independent from its nature (\textit{i.e.} aug-cc \textit{vs.} def2). Indeed, the difference between the two methods is divided by two with respect to the basis-set size. Anyway, due to the single reference nature of the ground state of polonium(IV) complexes, we indeed expect to obtain accurate values with MP2 provided that convergence of the results with respect to the basis-set size is reached. This is why we  chose to pursue of analysis of the results dependence with respect to the basis set with solely the MP2 method. 

Nowadays, computational facilities allow us to proceed with larger basis sets than fifteen years ago. For instance, we have reported calculations with the aug-cc-pVTZ, def2-TZVP and def2-QZVP basis sets (see Table~\ref{tbl:Eint}). By comparing the aug-cc-pVDZ and aug-cc-pVTZ results, it is clear that the MP2 interaction energies are not yet converged with aug-cc-pVDZ basis set, which somehow justifies \textit{a posteriori} the present benchmark. Thus, at least a triple-zeta basis set should be used. As an alternative to the aug-cc basis sets, we have employed the def2-TZVP and def2-QZVP ones (see Table \ref{tbl:Eint}). Notably, the def2-TZVP basis set show very similar results to the ones obtained with the aug-cc-pVTZ basis set, despite being built with fewer functions. The largest difference between the corresponding sets of results is below \SI{20}{\kcal} for the $n$ = 9 complex, which translates into $\sim$\SI{2}{\kcal}  per ligand. Since the calculations with aug-cc-pVTZ basis set take roughly 5 to 20 times more time than the def2-TZVP one, it is clear that the best accuracy/cost ratio is expected for the def2 basis sets, in particular the def2-TZVP one. 

As an additional step of analysis, we compared the results obtained with the def2-TZVP and def2-QZVP basis sets. As can be seen in Supporting Information (Tables S1 and S2), interaction energies differ by less than \SI{2}{\kcal} per ligand when the BSSE correction is included (the correction effectively reducing this difference). Considering the higher computational cost of calculations with the def2-QZVP basis set, we conclude that the MP2/def2-TZVP level of theory reaches the best accuracy to cost ratio out of all the tested methods for computing interaction energies between the \ce{Po^{4+}} ion and the coordinated water ligands. Furthermore, we do not expect a notable improvement of the interactions energies by performing a complete basis set extrapolation~\cite{karton2006comment,feller2008survey, feller2011effectiveness} based on the def2-TZVP and def2-QZVP calculations. 

\begin{figure}[ht]
\centering
  \includegraphics[width=0.9\linewidth]{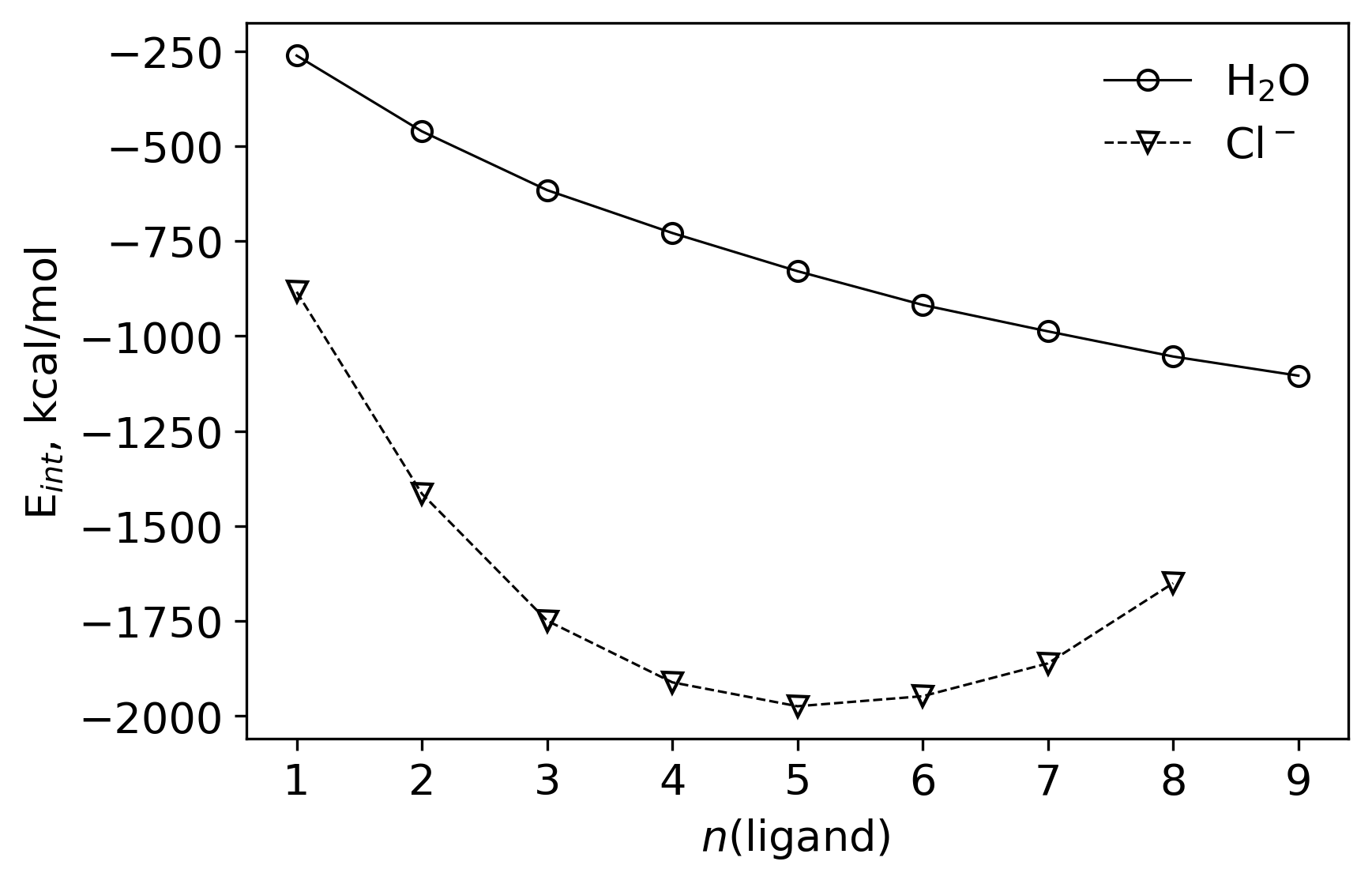}
  \caption{MP2/def2-TZVP interaction energies (\si{\kcal}) as functions of the number of ligands ($n$) for polonium(IV) complexes with water and chlorides. }
  \label{fgr:Eint}
\end{figure}

We have repeated similar calculations on the chlorinated complexes. First, we have found that eight chloride ligands are sufficient to saturate the coordination sphere of polonium(IV) ions. Second, we observe a change in the evolution of the interaction energies (see Table \ref{tbl:Eint}), which is \textit{a priori} more visible in Figure~\ref{fgr:Eint}.  While in the case of water, the evolution is monotonous with respect to ligand addition, the curve for chloride ligands displays a minimum for $n$ = 5 (note that all the structures are stable though). We connect this phenomenon with the negative charges of chloride ligands: for each ligand addition, we earn a coordination bond but we also enhance the inter-ligand repulsion in a much stronger way than with neutral ligands. We have also repeated a similar benchmark as what was done for the step-wise water addition (see Tables S3 and S4), leading to the exact same conclusion: the MP2/def2-TZVP level of theory is a good compromise to compute accurate interaction energies.

\begin{figure}
\centering
  \includegraphics[height=0.92\textheight]{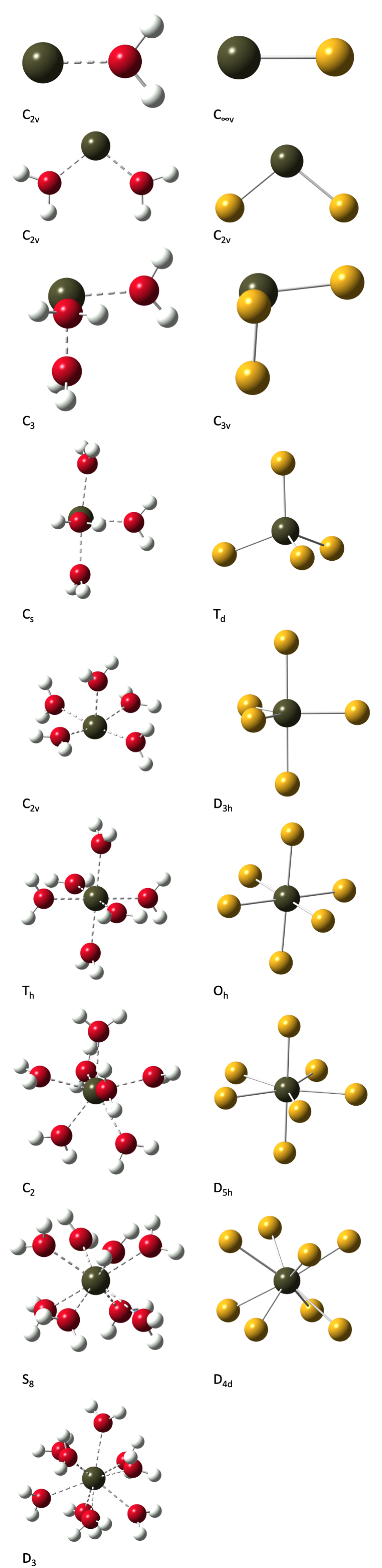}
  \caption{Ball-and-stick representations and symmetry point groups of the studied polonium(IV) complexes with water and chlorides. Colour code: polonium (gray), chlorine (yellow), oxygen (red) and hydrogen (white).}
  \label{fgr:geometries}
\end{figure}

As a part of our benchmark study, we have been looking at molecular geometries, with emphasis on the mean \ce{Po-O} and \ce{Po-Cl} bond lengths. Prior to fully digging into this, we start by an overall description of the structures. Perspective views are given in Figure~\ref{fgr:geometries}, together with the symmetry point groups (SPGs). For the polonium(IV) complexes with water, we obtain similar structures as the ones reported by Ayala~{\etal}~\cite{ayala2008po}, if we exclude the \ce{[Po(H2O)7]^{3-}} complex from the discussion (a different SPG may be found depending on the applied level of theory). Concerning this specific complex, note that we report a $C_2$ SPG, obtained at the MP2/def2-TZVP level of theory. 

When comparing the \ce{[Po(H2O)_n]^{4+}} and \ce{[PoCl_n]^{4-n}} structures two-by-two for $n=$ 1--8, several situations occurs:

\begin{itemize}
    \item For $n$ = 2, the same SPG is observed;
    \item For $n$ = 1, 3, 6 and 8, a direct subgroup of the SPG of the \ce{[PoCl_n]^{4-n}} complex is found for the corresponding \ce{[Po(H2O)_n]^{4+}} complex;
    \item For $n$ = 4, 5 and 7, the \ce{[Po(H2O)_n]^{4+}} complexes display, if any, much less symmetry elements than the corresponding \ce{[PoCl_n]^{4-n}} ones. Those structures are thus considered as qualitatively different.
\end{itemize}

\noindent Though it is obvious that the presence of the hydrogen atoms of the water ligands may lower the symmetry, it is actually interesting to note that in some of the cases ($n$ = 4, 5 and 7) quite different structures were obtained. 

Concerning more specifically the \ce{PoCl4} complex, Stoïanov \textit{et al.} reported in their article~\cite{stoianov2019uv} a $C_{2v}$ SPG for the gas-phase structure, with one set of bonds being slightly longer than the other. In fact, we found with our level of theory that the gas phase $C_{2v}$ structure is only \SI{1.6}{\kcal} higher in energy than the $T_d$ one, meaning that both are technically possible. We have thus performed several test calculations with various levels of theory, and concluded that though the potential energy hypersurface may be quite flat, the $T_d$ structure must be the correct gas-phase one (we recall here that a structure similar to the one of the corresponding water complex is expected when a solvation model is applied~\cite{stoianov2019uv}). Finally, we would like to mention that the structure of the \ce{[Po(H2O)9]^{4+}} complex, which has no match in the chlorinated complex series, is highly symmetric ($D_3$ SPG), with 6 and 3 by symmetry related bonds, respectively. 

\begin{figure}
\centering
  \includegraphics[width=0.9\linewidth]{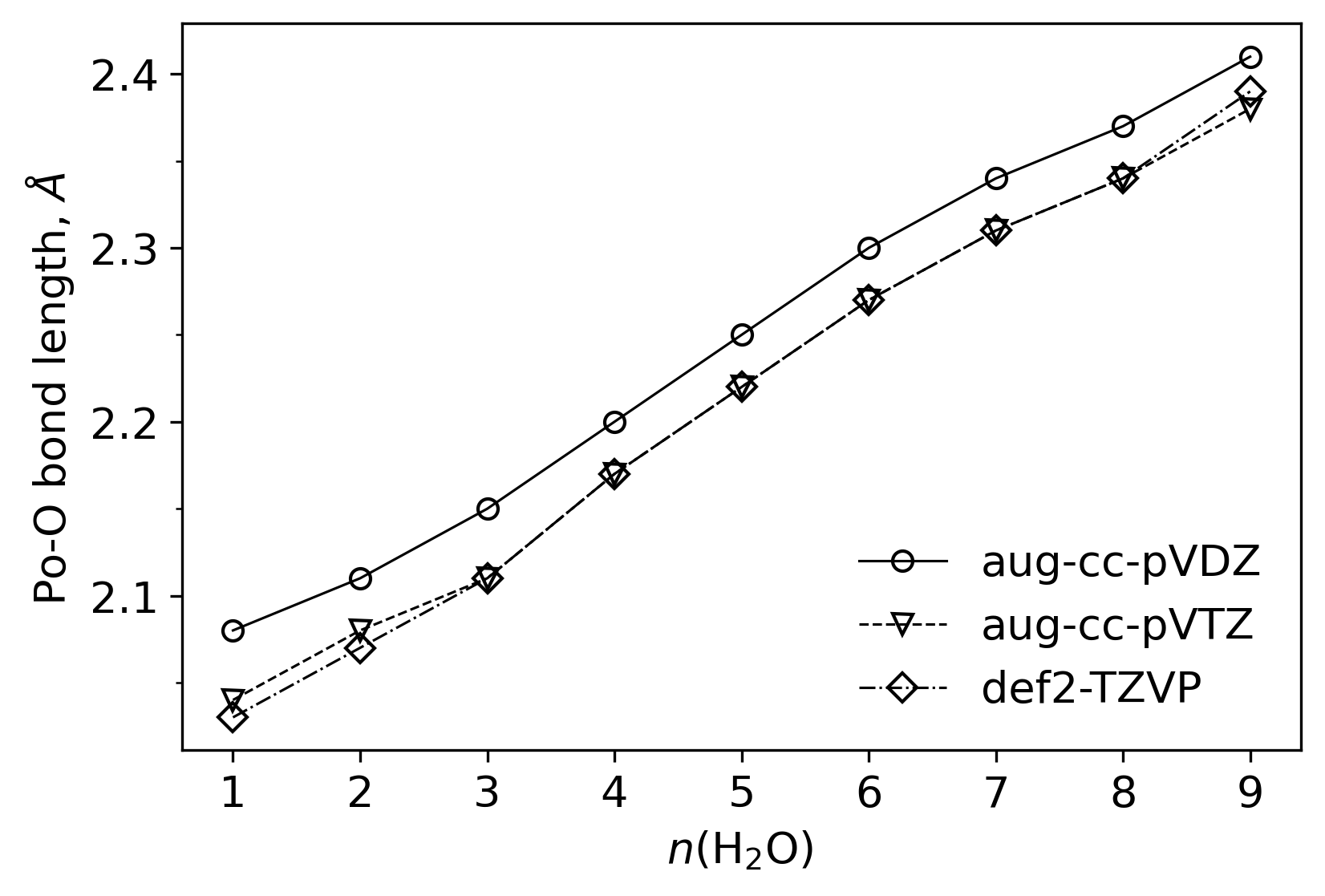}
  \caption{Mean MP2 \ce{Po-O} bond lengths (\si{angstrom}), obtained for polonium(IV) complexes with water for various basis sets, as functions of the number of ligands ($n$).}
  \label{fgr:waterbonds}
\end{figure}

The dependence of the mean \ce{Po-O} and \ce{Po-Cl} bond lengths with respect to the basis-set sizes may be seen from Tables \ref{tbl:bondwaterBS}, S5 and S6, and is also displayed in Figure \ref{fgr:waterbonds} for the polonium(IV) complexes with water. In line with the interaction energies, we note significant shortening of the bonds by going from the aug-cc-pVDZ basis set to the aug-cc-pVTZ one, a good agreement between the aug-cc-pVTZ and def2-TZVP results, and little change by going from def2-TZVP to def2-QZVP, meaning that the def2-TZVP basis set is again a good accuracy/cost ratio performer.

\begin{table}[ht]
	\small
	\caption{\ Mean MP2 \ce{Po-O} and \ce{PoCl} bond lengths (\si{\angstrom}), obtained for polonium(IV) complexes with water and chlorides, respectively}
	\label{tbl:bondwaterBS}
    \begin{tabular*}{0.48\textwidth}{@{\extracolsep{\fill}}l*{4}{S[table-format=1.2]}}
		\toprule
		& \multicolumn{3}{c}{H$_2$O} & {Cl-}\\
		\cmidrule{2-4}
		n(ligands)& {aug-cc-pVDZ} & {aug-cc-pVTZ} & {def2-TZVP} & {def2-TZVP}  \\ 
		\midrule
1 & 2.08 & 2.04 & 2.03 & 2.25 \\
2 & 2.11 & 2.08 & 2.07 & 2.28 \\
3 & 2.15 & 2.11 & 2.11 & 2.34 \\
4 & 2.20 & 2.17 & 2.17 & 2.47 \\
5 & 2.25 & 2.22 & 2.22 & 2.54 \\
6 & 2.30 & 2.27 & 2.27 & 2.60 \\
7 & 2.34 & 2.31 & 2.31 & 2.72 \\
8 & 2.37 & 2.34 & 2.34 & 2.83 \\
9 & 2.41 & 2.38 & 2.39 & \\ 
		\bottomrule
	\end{tabular*}
\end{table}

For the \ce{[Po(H2O)_n]^{4+}} ($n$ = 1--3) complexes, we obtain similar MP2/aug-cc-pVDZ bond distances than the ones of Ayala~{\etal}~\cite{ayala2008po}, meaning that our proposed change in the NFCs as virtually no effect on the molecular geometries. Consequently, the good agreement of the MP2 and CCSD(T) structures that they observed is preserved in the present study. Concerning the \ce{[PoCl_n]^{4-n}} ($n$ = 4--6) complexes, the mean B3LYP/aug-cc-pVTZ bond distances reported by Stoïanov~{\etal}~\cite{stoianov2019uv} (\SIlist{2.52; 2.59;2.67}{\angstrom}, respectively), are significantly longer than the MP2/def2-TZVP ones (by $\sim$\SI{0.05}{\angstrom}), which is in line with the $\sim$\SI{0.03}{\angstrom} observed by Ayala~{\etal}~\cite{ayala2008po} for the \ce{[Po(H2O)_n]^{4+}} ($n$ = 1--9) series. Note that the application of a dispersion correction~\cite{grimme2011effect} slightly diminishes those bond distances (by $\sim$\SI{0.01}{\angstrom} in the \ce{[PoCl_n]^{4-n}}, $n$ = 4--6, complexes), which thus only explains part of the difference between B3LYP and MP2. Since the bond lengths of the \ce{[Po(H2O)_n]^{4+}} (n = 1--3) complexes tend to be overestimated by $\sim$\SI{0.02}{\angstrom}~\cite{ayala2008po}, we conclude that our new MP2/def2-TZVP structures for the \ce{[PoCl_n]^{4-n}} ($n$ = 4--6) complexes are actually more accurate than the previously reported ones~\cite{stoianov2019uv}.    

\subsection{QTAIM analyses}

To better characterise the nature of the bonds in the studied polonium complexes, QTAIM~\cite{Bader:1991} analyses were performed. Several indicators may be defined, in particular at the bond critical points (BCPs). Such indicators may be used to classify the nature of the bonds, as proposed by Nakanishi and Hayashi~\cite{Nakanishi:2013a} and validated by Pilmé~{\etal}~\cite{pilme2014qtaim} for astatine species (a neighbour of polonium in the periodic table). We report for convenience the bond lengths, and more specifically the electron density values at the BCPs ($\rho$), the Laplacian of the density at the BCP ($\nabla^2\rho$), the kinetic energy density (G), the potential energy density (V), the ratio between the absolute value of the potential energy density to the kinetic energy one (|V|/G),  the delocalisation indices ($\delta$), and the Po and O or Cl atomic charges (q) in Tables ~\ref{tbl:QTAIM_Water} and \ref{tbl:QTAIM_Cl}. 

\begin{table*}[ht]
	\small
\caption{\ MP2/def2-TZVP \ce{Po-O} bond distances (\si{\angstrom}) and QTAIM bonding descriptors for polonium(IV) complexes with water}
\label{tbl:QTAIM_Water}
\begin{tabular*}{\textwidth}{@{\extracolsep{\fill}}
l 
S[table-format=1.2] 
l 
*{3}{S[table-format=1.2]} 
S[table-format=3.2] 
*{2}{S[table-format=1.1]} 
*{3}{S[table-format=3.2]}
}
	\toprule
		$n$(\ce{H2O}) & \multicolumn{2}{c}{$r$(\ce{Po-O}), \si{\angstrom}} & {$\rho$, a.u.} & {$\nabla^2\rho$} & {G, a.u.} & {V, a.u.}  & {|V|/G} & {$\delta$(\ce{Po-O})} & {q(Po), a.u.}& {q(O) , a.u.}& {q(H) , a.u.}\\
		\midrule
	1 & 2.03 &     & 0.20 & 0.42 & 0.15 & -0.20 & 1.30 & 1.1 & 3.36 & -1.04 & 0.84 \\
		2 & 2.07 &     & 0.12 & 0.36 & 0.13 & -0.17 & 1.31 & 0.8 & 3.03 & -1.13 & 0.80 \\
		3 & 2.11 &     & 0.11 & 0.31 & 0.11 & -0.15 & 1.31 & 0.6 & 2.91 & -1.20 & 0.78 \\
		4 & 2.13 &     & 0.10 & 0.29 & 0.10 & -0.14 & 1.31 & 0.5 & 2.87 & -1.23 & 0.77 \\
		& 2.22 &(x2) & 0.08 & 0.24 & 0.08 & -0.11 & 1.28 & 0.6 &  & -1.25 &  \\
		& 2.12 &     & 0.10 & 0.30 & 0.11 & -0.14 & 1.31 & 0.6 &  & -1.21 &  \\
		5 & 2.13 &     & 0.10 & 0.30 & 0.10 & -0.13 & 1.29 & 0.4 & 2.85 & -1.23 & 0.75 \\
		& 2.24 &(x2) & 0.08 & 0.22 & 0.08 & -0.10 & 1.27 & 0.5 &  & -1.26 & 0.74 \\
		& 2.24 &(x2) & 0.08 & 0.22 & 0.08 & -0.10 & 1.27 & 0.5 &  & -1.26 & 0.74 \\
		6 & 2.27 &     & 0.07 & 0.21 & 0.07 & -0.09 & 1.26 & 0.4 & 2.84 & -1.26 & 0.73 \\
		7 & 2.32 &(x2) & 0.07 & 0.18 & 0.06 & -0.07 & 1.23 & 0.3 & 2.88 & -1.27 & 0.71 \\
		& 2.34 &(x2) & 0.06 & 0.18 & 0.06 & -0.07 & 1.23 & 0.3 &  & -1.27 & 0.72 \\
		& 2.29 &     & 0.07 & 0.20 & 0.06 & -0.08 & 1.24 & 0.3 &  & -1.26 &  \\
		& 2.28 &(x2) & 0.07 & 0.20 & 0.07 & -0.08 & 1.24 & 0.4 &  & -1.26 &  \\
		8 & 2.34 &     & 0.06 & 0.18 & 0.06 & -0.07 & 1.21 & 0.3 & 2.90 & -1.27 & 0.70 \\
		9 & 2.43 & (x3) & 0.05 & 0.15 & 0.04 & -0.05 & 1.17 & 0.2 & 2.92 & -1.26 & 0.69 \\
		\bottomrule
	\end{tabular*}
\end{table*}

\begin{table*}[ht]
	\small
	\caption{\ MP2/def2-TZVP \ce{Po-Cl} bond distances (\si{\angstrom}) and QTAIM bonding descriptors for polonium(IV) complexes with chlorides}
	\label{tbl:QTAIM_Cl}
\begin{tabular*}{\textwidth}{@{\extracolsep{\fill}}
l 
S[table-format=1.2] 
l 
*{3}{S[table-format=1.2]} 
S[table-format=3.2] 
*{2}{S[table-format=1.1]} 
*{2}{S[table-format=3.2]}
}
		\toprule
		{$n$(\ce{Cl-})} & \multicolumn{2}{c}{$r$(\ce{Po-Cl}), \si{\angstrom}} & {$\rho$, a.u.} & {$\nabla^2\rho$} & {G, a.u.} & {V, a.u.} &  {|V|/G} &  {$\delta$(\ce{Po-Cl})} & {q(Po) , a.u.}& {q(Cl) , a.u.}\\ 
		\midrule
		1 & 2.25 & & 0.12 & 0.26 & 0.12 & -0.17 & 1.46 & 1.8 & 2.52 & 0.48 \\
		2 & 2.28 & & 0.12 & 0.18 & 0.10 & -0.15 & 1.53 & 1.2 & 2.06 & -0.03 \\
		3 & 2.34 & & 0.11 & 0.13 & 0.08 & -0.12 & 1.57 & 0.9 & 1.94 & -0.31 \\
		4 & 2.47 & & 0.08 & 0.13 & 0.06 & -0.09 & 1.46 & 0.7 & 1.98 & -0.50 \\
		5 & 2.56 & (x2) & 0.07 & 0.11 & 0.05 & -0.07 & 1.43 & 0.6 & 2.06 & -0.64 \\
		& 2.53 & (x3) & 0.07 & 0.11 & 0.05 & -0.07 & 1.44 & 0.6 &  & -0.59 \\
		6 & 2.60 & & 0.06 & 0.10 & 0.04 & -0.06 & 1.41 & 0.5 & 2.13 & -0.69 \\
		7 & 2.75 & (x5) & 0.05 & 0.07 & 0.03 & -0.04 & 1.34 & 0.4 & 2.13 & -0.74 \\
		& 2.63 & (x2) & 0.06 & 0.09 & 0.04 & -0.05 & 1.39 & 0.5 &  & -0.71 \\
		8 & 2.83 & & 0.04 & 0.07 & 0.02 & -0.03 & 1.28 & 0.4 & 2.12 & -0.77 \\
		\bottomrule
	\end{tabular*}
\end{table*}

With the value of the electron density at the BCP of interest, one can determine the nature of the bond, in particular, determine if the interaction is of the \emph{shared-shell} ($\rho\geq0.2$ a.u.) or \emph{closed-shell} type. This classification may also be done based on the following two relations for the Laplacian ($\nabla^2\rho(\pmb{r})$) and the total electron energy densities at BCPs ($H(\pmb{r}$))~\cite{Nakanishi:2013a}:
\begin{align}
  \frac{1}{4} \nabla^2\rho(\pmb{r}) &= 2G(\pmb{r})\ +\ V(\pmb{r})\label{Eq:BCPLaplace}\\
  H(\pmb{r}) &= G(\pmb{r})\ +\ V(\pmb{r})\label{Eq:BCPH}
\end{align} 
where $G(\pmb{r})$ is positive and $V(\pmb{r})$ is negative. Actually, if the Laplacian of the density at the BCP is positive, as in all the studied complexes, then the kinetic energy density dominates, meaning that we are dealing with a \emph{closed-shell} interaction~\cite{Nakanishi:2013a}. 

Another criterion that can help us with the description of chemical bonds is the ratio between the absolute value of the potential energy density to the kinetic energy one. If this ratio is smaller than 1, we expect a \emph{pure closed-shell} interaction (thus a pure ionic bond or a typical non-covalent interaction). If this ratio is larger than 1, which is the case in all the studied complexes, it means that the potential energy is large (electrons are thus stabilized at the BCP~\cite{Nakanishi:2013a}), we face a \emph{regular closed-shell} interaction.

For both the types of complexes we can see the same trend: the larger the complex, the longer the bond, the smaller the values of electron density at the BCP, and so do the values of the Lagrangian, potential and kinetic energy densities. This is somehow logical since the addition of ligand will lead to smaller charge transfers from each ligand to the polonium(IV) ion, \textit{i.e.} to reduced ``covalency'', and since inter-ligand repulsion is also enhanced (which obviously is much more pronounced for chloride ligands, as already mentioned). 

\begin{figure}[ht]
\centering
  \includegraphics[width=0.9\linewidth]{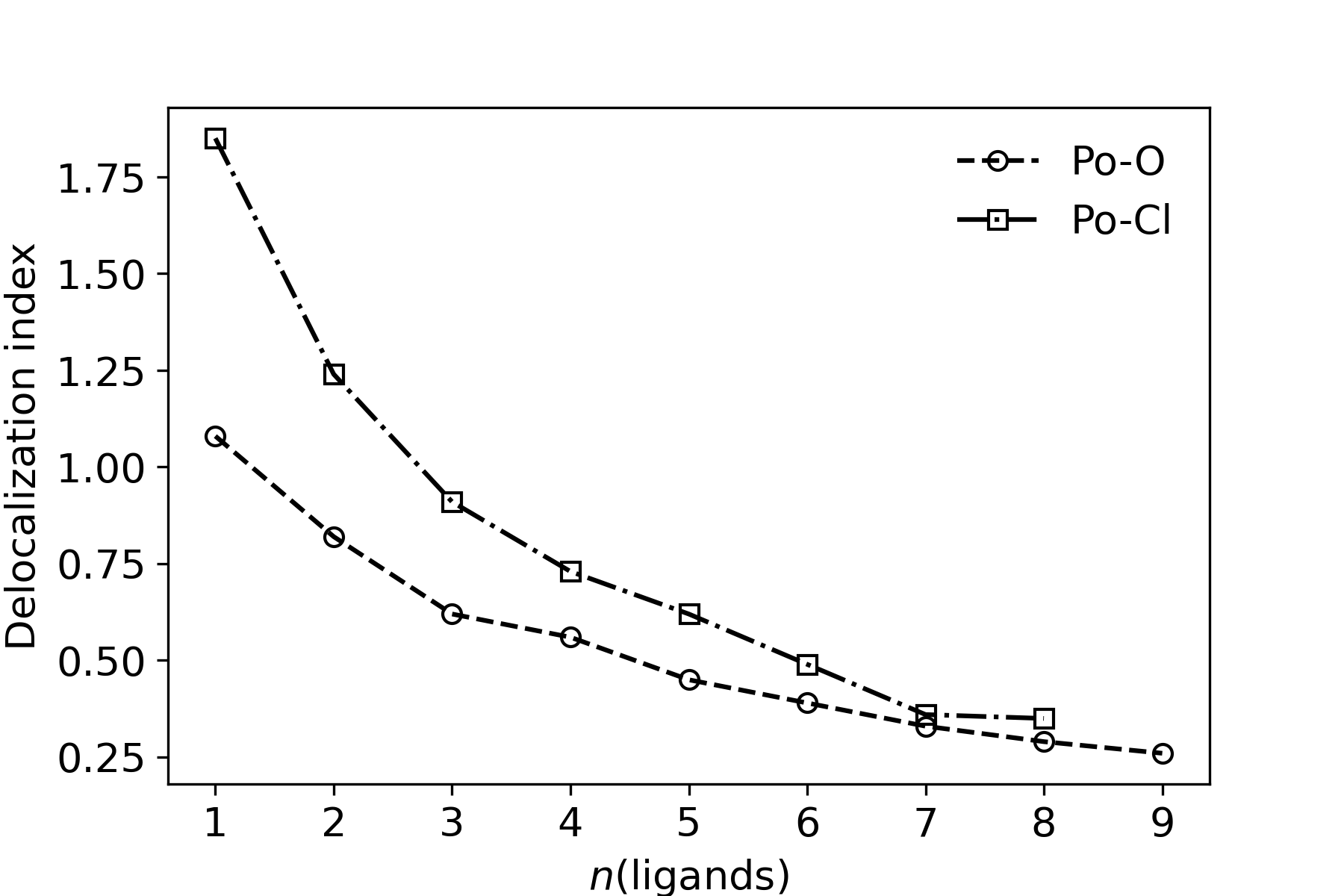}
  \caption{Mean MP2/def2-TZVP \ce{Po-O} and \ce{Po-Cl} delocalisation indices as functions of the number of ligands ($n$) for polonium(IV) complexes with water and chlorides, respectively.}
  \label{fgr:DI}
\end{figure}

The delocalisation index ($\delta$)~\cite{Bader:1975, Fradera:1999}, is also another important criterion, meant to be related to the bond multiplicity. A plot of the comparison between the mean $\delta$(\ce{Po-O}) and $\delta$(\ce{Po-Cl}) values is depicted in Figure \ref{fgr:DI}, allowing a quicker view at the general trends than by inspecting Tables \ref{tbl:QTAIM_Water} and \ref{tbl:QTAIM_Cl}. First, we can see that almost all the values are lower or equal to 1, that represents an ideal  single bond (the two atoms exactly sharing one electron pair). In fact, half of the complexes presented here have $\delta$'s that are smaller than 0.5, meaning that the corresponding bonds are far from being covalent. Since a $\delta$ value around 1 may be indicative of several bonding patterns (\textit{i.e.} covalent, charge-shift~\cite{Shaik:2009} or even iono-covalent), these are not incompatible with our previous conclusion (\emph{regular closed-shell} interactions). In fact, the more striking value is the one observed for the \ce{[PoCl]^{3+}} complex: 1.8. This value somehow resembles what is expected for a double bond, meaning that this bond must bear a strong covalent character. 

In line with the delocalisation index, the QTAIM atomic q(Cl) charge in \ce{[PoCl]^{3+}} is positive (strong \ce{Cl-} to \ce{Po^{4+}} charge transfer). Knowing that atomic charges are strongly dependent on the underlying charge model, we have also computed the Mulliken~\cite{Mulliken:1955} and Hirschfeld~\cite{Hirshfeld:1977} atomic charges (see Table S7). Since all the applied charge models lead to a positive q(Cl) in \ce{[PoCl]^{3+}}, we conclude for a strong charge transfer from \ce{Cl-} to \ce{Po^{4+}} in this diatomic. The trends observed by increasing the number of ligand is identical for the three charge models and is in fact not surprising: q(Cl) always decrease with increasing $n$. 

\subsection{Spin-orbit coupling calculations}

As already mentioned, the free \ce{Po^{4+}} ion has a [Xe]4f$^{14}$5d$^{10}$6s$^2$ electronic structure. We recall here that within the non-relativistic or scalar-relativistic picture (\textit{i.e.} when one considers orbitals and not spinors), the SOC is only active with the occurrence of partially filled p, d, f, \textit{etc.} shells (note that this is a necessary but not sufficient condition). Therefore, electron donation to the \ce{Po^{4+}} ion is required to activate the SOC, and in fact we have just seen that such a strong charge transfer occurs for the \ce{[PoCl]^{3+}} complex, which encouraged us to analyse the role of the SOC on this \ce{Po-Cl} bond, in particular concerning the bond distance. 

\begin{figure}[ht]
\centering
  \includegraphics[width=0.9\linewidth]{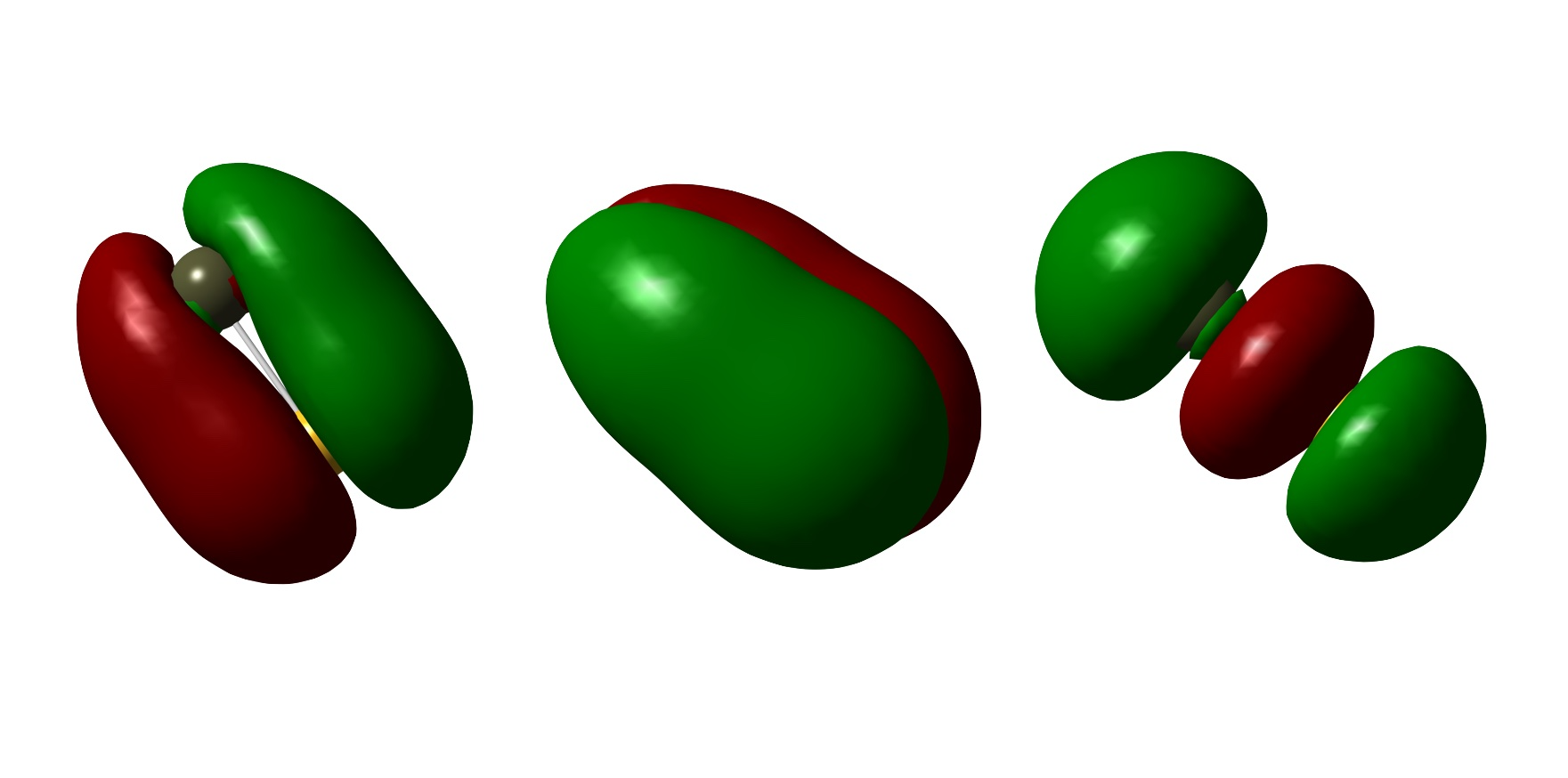}
  \caption{Representations of the HOMO, HOMO$-$1 and HOMO$-$2 $\pi$ and $\sigma$ bonding orbitals of the \ce{[PoCl]^{3+}} complex. Note that HOMO and HOMO$-$1 are degenerate by symmetry (those are the $\pi$ orbitals).}
  \label{fgr:PoCl_orb}
\end{figure}

At the non-relativistic or scalar-relativistic Hartree-Fock level, three bonding valence orbitals are doubly occupied and have no populated antibonding counterparts (see Figure \ref{fgr:PoCl_orb}). These are the doubly degenerate $\pi$-bonding HOMO and HOMO-1 orbitals and the HOMO-2 $\sigma$-bonding orbital. This $\sigma$-bonding orbital is lower in energy than the $\pi$-bonding ones, meaning that the molecular orbital diagram is \textit{uncorrelated}. Below these, the two anti bonding and bonding $\sigma$ orbitals are essentially formed by linear combinations of the 6s orbital of Po and of the 3s orbital of Cl. Those two extra valence orbitals are doubly occupied, giving a net zero contribution to the formal bond order. Overall, the formal bond order in this system is 3, and the populated bonding orbitals are (i) delocalised on the two atoms (see Figure \ref{fgr:PoCl_orb}) and (ii) essentially formed by linear combinations of the 6p orbitals of Po and of the 3p orbitals of Cl. 

Without explicitly computing effective bond orders (EBOs)~\cite{Roos:2007,Gendron:2014a,Maurice:2015a,Knecht:2019,GomezPech:2019}, we would like to stress that two effects will tend to weaken the bond and thus enlarge the bond distance:

\begin{itemize}
    \item electron correlation may promote electron pairs from a bonding to an antibonding orbital, here especially from the $\sigma$ orbital to the $\sigma$* one (correlation induced reduction of the $\sigma$ contribution to the EBO);
    \item the SOC may promote single electrons to antibonding orbitals, here especially the $\pi$ orbitals to the $\sigma$* one (SOC induced reduction of both the $\pi$ and $\sigma$ contributions to the EBO).
\end{itemize}

\noindent As a consequence, the effective bond order will significantly differ from the formal one, meaning that one should not see the \ce{Po-Cl} bond in \ce{[PoCl]^{3+}} as triple (and we have already seen with the delocalisation index, in the absence of the SOC, is more likely to be a double bond). 

In previous works~\cite{ayala2008po, stoianov2019uv}, the role of the SOC on the molecular geometries of polonium(IV) complexes was neglected, based on reference to the electronic structure of the free \ce{Po^{4+}} ion. However, since we have here evidenced that a strong effect can potentially occur in the hypothetical \ce{[PoCl]^{3+}} complex, we have computed molecular geometries in the absence and in the presence of the SOC, with two different frameworks, which are in fact complementary (see Computational details section):

\begin{itemize}
    \item ZORA scalar relativistic and two-component Hartree-Fock calculations (no electron correlation, \textit{a priori} inclusion of the SOC);
    \item two-step SOCI calculations, with a first scalar relativistic step and then diagonalisation of the  \mbox{\textbf{H}$_\text{tot}$ = \textbf{E}$_\text{el}$ + \textbf{H}$_\text{SOC}$} matrix in the basis of the $M_S$ components of the states of the previous step (spin-orbit free electron correlation, \textit{a posteriori} inclusion of the SOC).
\end{itemize}

\noindent We just recall here that the first approach may overestimate the impact of the SOC (by lacking correlation) and that the second one (contracted SOCI) may underestimate it (by neglecting the SOC \textit{polarisation}~\cite{Vallet:2000}).

First, we tried out the \emph{a priori} inclusion of the SOC, using the AMS software. For the  \ce{[PoCl]^{3+}} complex, we observed an increase in the bond length from \SIrange{2.17}{2.29}{\angstrom}, \textit{i.e.} a \SI{0.12}{\angstrom} SOC-induced bond lengthening. This is a significant effect, clearly larger than the expected error of the underlying quantum mechanical method (meaning that this effect is significant and thus not negligible). As we have seen that ``covalency'', a key ingredient to active the SOC in polonium(IV) complexes, quickly diminishes in the \ce{[PoCl_n]^{4-n}} series, we have performed calculations for the $n$ = 2, 3, 4 and 8 complexes. For the \ce{[PoCl2]^{2-}} complex, the SOC triggers a \SI{0.01}{\angstrom} bond lengthening, which becomes even more insignificant for larger $n$ values. From this first SOC study, we can already conclude that the SOC is negligible in high-coordinated polonium(IV) complexes, implying that its inclusion should not be key to perform future calculations of polonium(IV) complexes in solution (solvent molecules will always surround the heavy cation). On the contrary, a significant SOC effect can potentially occur on low-coordinated polonium(IV) complexes, meaning that SOC should somehow generally be probed in diatomics or systems with only a coupled of coordinated ligands (those systems are mostly of theoretical interest since gas phase experiments with polonium are not that safety friendly).

Alternatively, we have also performed SOCI calculations on the \ce{[PoCl]^{3+}} complex, which proved to be interesting in terms of the SOC. We recall here that the two-step SOCI implementations do not yet propose any numerical or analytical gradients, meaning that the user has to explore the potential energy surfaces, here the potential energy curve (obviously with respect to the bond distance), by manual scans~\cite{Sergentu:2015}. Several degrees of freedom are inherent to the SOCI approach: the size of the active space in the state-average complete active space (SA-CASSCF) calculation, the number of states that are included in the SA-CASSCF calculation and the way post-CASSCF electron correlation is introduced. In our calculations, we have considered 6 active electrons within 6 active orbitals (the $\sigma$, $\pi$, $\pi$* and $\sigma$* orbitals that are essentially formed by linear combinations of the 6p orbitals of Po and of the 3p orbitals of Cl). Note that test calculations showed that enlargement of the active space hardly affected our conclusions. With a ground state spin-singlet state, only spin-triplet spin components are directly coupled to it. After some tests, we have set up the SA-CASSCF state space to 11 spin-triplet and 1 spin-singlet states. Post-CASSCF electron correlation has been introduced by replacing the \textbf{E}$_\text{el}$ diagonal elements of the SOCI matrix~\cite{Teichtel:1983, Llusar:1996} by $N$-electron valence space perturbation theory at second order (NEVPT2) ones~\cite{Dyall:1995, Angeli:2001a, Angeli:2001b}. With this setup, we observed a SOC-induced bond lengthening of \SI{0.08}{\angstrom} (from \SIrange{2.27}{2.35}{\angstrom}). 

Before concluding on the role of the SOC on the bond distance, it is important to comment on the difference between both the scalar relativistic bond distances reported above (\SI{2.17}{\angstrom} \textit{vs.} \SI{2.27}{\angstrom}). For the sake of pedagogy and completeness, we would like to mention that three main factors may affect this difference: the basis set (TZ2P \textit{vs.} a combination of DKH-def2-TZVP and SARC-DKH-TZVP, depending on the atom), the scalar relativistic Hamiltonian (ZORA \textit{vs.} Douglas-Kroll-Hess) and the reference electronic structure method (Hartree-Fock \textit{vs.} NEVPT2). As mentioned before, electron correlation is meant to weaken the bond order and in fact, also enlarge the bond distance. To probe this, we have performed an additional Hartree-Fock scan with ORCA with the same setup as in the second set of calculations, reproducing an equilibrium bond distance of \SI{2.17}{\angstrom}, as in the first set of calculations. We thus essentially attribute the observed difference in scalar relativistic bond distances to electron correlation. 

Since the second approach is expected to underestimate the effect of SOC on the bond distance while the first one should overestimate it (see above), we finally conclude for a $\sim$\SI{0.1}{\angstrom} SOC-induced bond lengthening in \ce{[PoCl]^{3+}}. This lengthening is of the same order of magnitude as the one observed for the AtX (X = F--At) diatomic molecule for instance~\cite{GomezPech:2019}, for which the importance of the SOC was more obvious (the free At atom displaying a strong SOC splitting of $\sim$\SI{2.9}{\electronvolt}~\cite{Maurice:2015a} in its ground $^2$P state). Also, it was shown in these systems that the SOC reduces the covalent character of the bond in terms of EBO~\cite{GomezPech:2019} and of topological indicators~\cite{Pilme:2012,pilme2014qtaim,GomezPech:2020} such as the delocalisation index or the population of the bonding basin of the electron localization function~\cite{Becke:1990,Silvi:1994, Savin:1996}.

\section{Conclusions}

In this work, two series of polonium(IV) complexes were investigated, respectively with water and chloride ligands. A new benchmark study performed on the water complexes showed that the MP2/def2-TZVP level of theory is both accurate and efficient. 

We have seen that the coordination sphere of the polonium(IV) ion may fit up to 9 water molecules or 8 chloride ligands, respectively. The interaction energy curves built by increasing $n$ do not follow the exact same trend, a fact which we have attributed to the neutral \textit{vs.} charged nature of the ligands.

We have also analysed the chemical bonds in the studied systems, generally qualified as \emph{regular closed-shell} interactions. Because of signs of a strong ``covalency'' in the \ce{[PoCl]^{3+}} complex, we have investigated the influence of the SOC on the \ce{Po-Cl} bond distance, quite unexpectedly revealing a SOC-induced bond lengthening of $\sim$\SI{0.1}{\angstrom} in this system, which vanishes for larger coordinations (in fact already starting from $n$ = 2). This allows us to formulate a more general conclusion concerning heavy-element chemistry:

\begin{itemize}
    \item[] Even if the reference free ion of a given heavy-element cation do not display any SOC, electron donation from the ligands to the cation may be strong enough to active it and trigger sizeable effects. As a consequence, computational chemists should definitely not neglect the role of the SOC \textit{a priori} (\textit{i.e.} without any testing), especially in the case of low-coordinated complexes.
\end{itemize}

Two main perspectives arise from this work, (i) exploring more deeply and largely the role of the SOC in low-coordinated polonium(IV) complexes as well as with valence iso-electronic complexes such as bismuth(III) ones and (ii) studying thoroughly the role of solvation on the molecular geometries, interaction energies and bonding in those complexes, which was out of the scope of the present work.

\section*{Author Contributions}
All the calculations were performed by N.Z. All the authors contributed to the scientific development of this work and jointly discussed the interpretation of the results. N.Z. wrote the first draft of the manuscript that has been then amended by all the co-authors.

\section*{Conflicts of interest}
There are no conflicts of interest to declare.

\section*{Acknowledgements}
This work has been supported by the MITI of CNRS (80|Prime project MSM4Po). HPC resources from the CCIPL (``Centre de calcul intensif des Pays de la Loire'') have been used. FR and VV acknowledge support from PIA ANR project CaPPA (ANR-11-LABX-0005-01), I-SITE ULNE projects OVERSEE and MESONM International Associated Laboratory (LAI) (ANR-16-IDEX-0004), and the French Ministry of Higher Education and Research, region Hauts de France council and European Regional Development Fund (ERDF) project CPER CLIMIBIO.



\balance


\bibliography{PoComplexes} 
\bibliographystyle{rsc} 

\end{document}